\documentclass[8.5pt,twoside,twocolumn]{article}
\usepackage[super,sort&compress,comma]{natbib}
\usepackage{mhchem}
\usepackage{times,mathptm}
\usepackage{sectsty}
\usepackage{balance}
\usepackage{color}

\usepackage{graphicx} 
\usepackage{lastpage}
\usepackage[format=plain,justification=raggedright,singlelinecheck=false,font=small,labelfont=bf,labelsep=space]{caption}
\usepackage{fancyhdr}
\begin{document}

\thispagestyle{plain}
\fancypagestyle{plain}{
\renewcommand{\headrulewidth}{1pt}}
\renewcommand{\thefootnote}{\fnsymbol{footnote}}
\renewcommand\footnoterule{\vspace*{1pt}%
\hrule width 3.4in height 0.4pt \vspace*{5pt}}

\makeatletter
\renewcommand\@biblabel[1]{#1}
\renewcommand\@makefntext[1]%
{\noindent\makebox[0pt][r]{\@thefnmark\,}#1}
\makeatother
\renewcommand{\figurename}{\small{Fig.}~}
\sectionfont{\large}
\subsectionfont{\normalsize}

\fancyfoot{}
\fancyhead{}
\renewcommand{\headrulewidth}{1pt}
\renewcommand{\footrulewidth}{1pt}
\setlength{\arrayrulewidth}{1pt}
\setlength{\columnsep}{6.5mm}
\setlength\bibsep{1pt}

\twocolumn[
  \begin{@twocolumnfalse}
\noindent\LARGE{\textbf{Following in the footsteps of {\it E. coli}: sperm in microfluidic ``strictures"}}
\vspace{0.6cm}

\noindent\large{\textbf{ E. Altshuler$^{1,2}$, G. Mi{\~n}o$^{3,2}$, A. Lindner$^2$, A. Rousselet$^2$, and E. Cl\'{e}ment$^{2}$}}\vspace{0.5cm}

\vspace{0.5cm}
 \end{@twocolumnfalse}
  ]

\footnotetext{$^1$ Group of Complex Systems and Statistical Physics, Physics Faculty, University of Havana, 10400 Havana, Cuba
\\ $^2$ PMMH, UMR 7636, CNRS, ESPCI Paris, PSL Research
University, Universit\`{e} Paris Diderot, Sorbonne Universit\`{e}, Paris, 75005, France
\\$^3$ LAMAE, Facultad de Ingenier\'{\i}a, Universidad Nacional de Entre R\'{\i}os (FI - UNER) and
Instituto de Investigaci\'{o}n y Desarrollo en Bioingenier\'{\i}a y Bioinform\'{a}tica (IBB) - CONICET - UNER, Argentina}

In a very interesting paper, Zaferani {\it et al.} study the locomotion of sperm into ``strictures" made on microfluidic channels that resemble an hourglass through which a liquid flow is established \cite{Zaferani2019}. Their experiments are, in fact, extremely similar to those published by Altshuler {\it et al.} in 2013 on another micro-swimmer, {\it E. coli} bacteria \cite{Altshuler2013}.

Figure 1 shows the clear parallel between the findings for bacteria and sperm in terms of motion: both tend to move against the flow along lateral boundaries; both detach and re-attach to lateral boundaries, depending on the relation between the local shear and the ``strength" of the swimmer. {\it But it should not surprise us}. Firstly, because {\it E. coli} and sperm share a lot of similarities. Both are ``pushers", i.e., their hydrodynamic center lies near the “head” of the swimmer, defined with respect to its direction of self-propelled motion and the resulting hydrodynamic interactions with bounding walls lead in both cases to an attraction \cite{ShunPak2015}. In addition, both swimmer shapes show a fore-aft asymmetry, which is thought responsible for the upstream motion observed for both micro-swimmers \cite{Tung2015, Kanstler2014, Kaya2012, Hill2007}. Secondly, because the geometry and dimensions of the ``hourglass" micro-channel structures used in \cite{Zaferani2019} are very similar to the ones used in \cite{Altshuler2013}.

\begin{figure}[!ht]
\centerline{\includegraphics[width=0.4\textwidth]{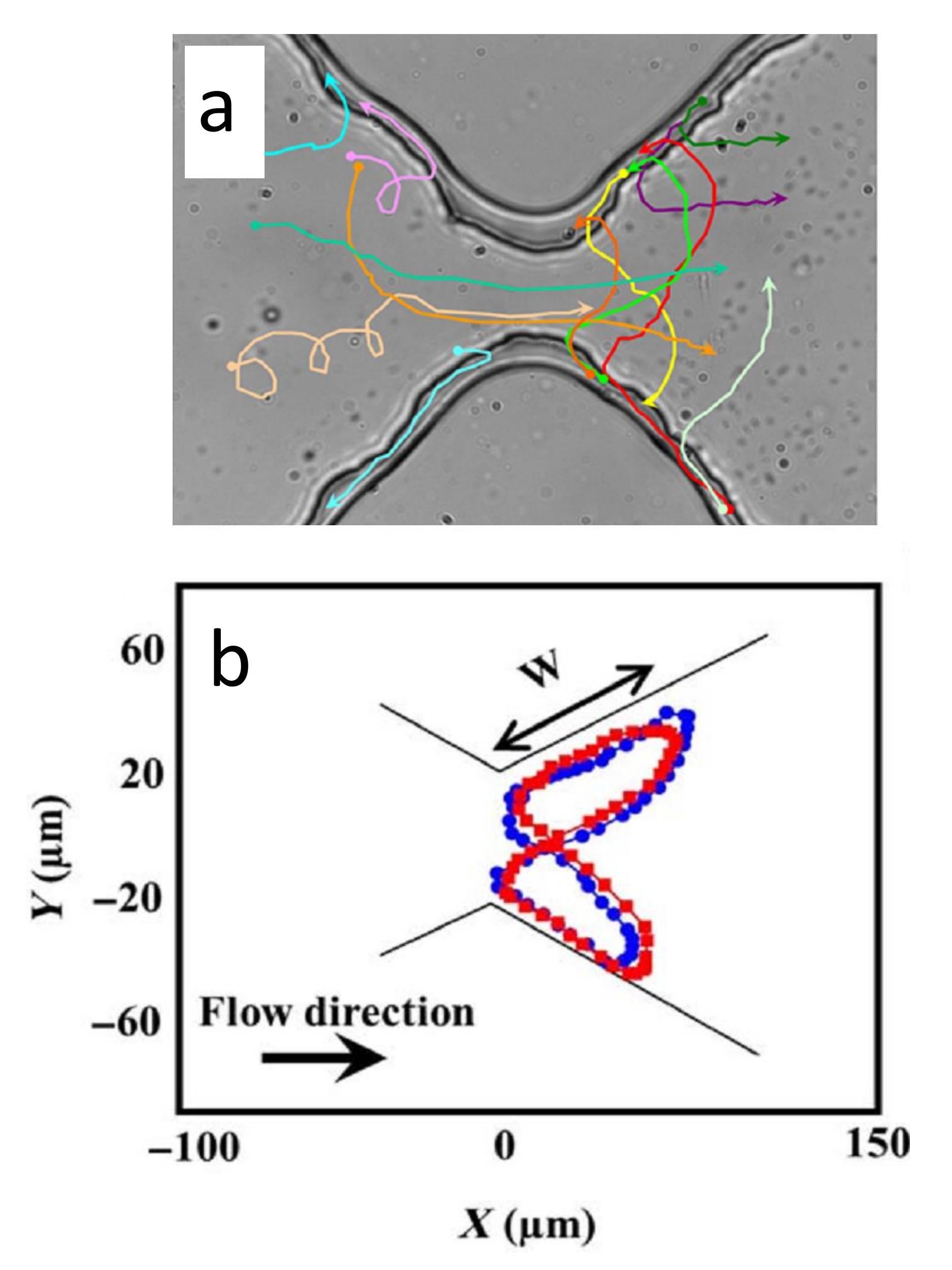}}
\caption{Following the steps of {\it E. coli} into a microfluidic ``stricture".(a) Figure 5(c) taken from the $2013$ paper by Altshuler {\it et al.} \cite{Altshuler2013} where the trajectories of {\it E. coli} bacteria are traced into an hourglass constriction made on a microfluidic channel where the liquid flow is established from left to right. The stricture is 40 $\mu$m width and forms 90 degrees angle; channel's width and depth are 200 $\mu$m and 20 $\mu$m, respectively. (b) Figure 3(c) taken from the $2019$ paper by Zaferani {\it et al.} \cite{Zaferani2019} where a similar experiment is performed on sperm. The stricture is of 40 $\mu$m width and forms 80 degrees angle; channel's width and depth are 300 $\mu$m and 30 $\mu$m, respectively. Key features of the locomotion of the micro-swimmers are shared between the two pictures (see text).}
\label{Figure1}
\end{figure}

Interestingly, equation (1) proposed by Zaferani {\it et al.} can be assumed as a ``microscopic" version of the advection-diffusion equation proposed by Altshuler {\it et al.} \cite{Altshuler2013}:

\begin{equation}\label{AdvDifGeneral}
-D \frac{d^2n(x)}{dx^2}+ \frac{d}{dx}[u_{a}(x) n(x)]=S(x)
\end{equation}

where $n(x)$ is the volume concentration of bacteria, $u_{a}(x)$ is their mean advection velocity along
the flow, $D$ is an effective longitudinal dispersion coefficient and $S(x)$ is a conservative bulk
source/sink term coming from the lateral wall contributions (i.e., absorption-erosion processes). By extracting from the experiment
the effect of the lateral walls given by $S(x)$, Altshuler {\it et al.} used the equation above to reproduce quantitatively the bacterial distribution
not only near the constriction, but far from it, as illustrated in Fig. 3 in \cite{Altshuler2013}.

The PMMH-ESPCI group's more recent work reveals the complex nature of $S(x)$ through systematic experiments on {\it E. coli} bacteria confined into
microfluidic channels at different flows \cite{Figueroa2015} --a very relevant study in connection to hospital infections. We believe that many features of the swimmer-boundary interaction in the presence of shear found in those experiments are valid to many pushers, so they have good chances to be extended for the important case of sperm.


\begin{thebibliography}{10}
%
\bibitem{Zaferani2019} M. Zaferani, G. D. Palermo and A. Abbaspourrad, Sci. Adv. \textbf{5}, eaav2111 (2019).
\bibitem{Altshuler2013} E. Altshuler, G. Mi\~{n}o , C. P\'{e}rez-Penichet, L. del R\'{\i}o, A. Lindner, A. Rousselet and E. Cl\'{e}ment, Soft Matter \textbf{9}, 1864 (2013)
\bibitem{ShunPak2015} Shun Pak and E. Lauga, Theoretical models in low-Reynolds number
locomotion In Fluid-Structure Interactions in Low- Reynolds-Number Flows (ed. C. Duprat and H. A. Stone), Royal Society of Chemistry.
\bibitem{Kanstler2014} V. Kantsler, J. Dunkel, M. Blayney, and R. E. Goldstein,
eLife \textbf{3}, 02403 (2014).
\bibitem{Tung2015} C. K. Tung, F. Ardon, A. Roy, D. L. Koch, S. S. Suarez, and M. Wu, Phys. Rev. Lett. \textbf{114}, 108102 (2015).
\bibitem{Kaya2012} T. Kaya and H. Koser, Biophys. J. \textbf{102}, 1514 (2012).
\bibitem{Hill2007} J. Hill, O. Kalkanci, J. McMurry, and H. Koser, Phys.Rev. Lett. \textbf{98}, 068101 (2007).
\bibitem{Figueroa2015} N. Figueroa-Morales, G. Mi\~{n}o, A. Rivera, R. Caballero,  E. Cl\'{e}ment, E. Altshuler and A. Lindner, Soft Matter \textbf{11}, 6284 (2015)

\end{thebibliography}
\end{document}